\documentclass[12pt]{article}
\usepackage{amsfonts}

\usepackage{amsfonts}
\usepackage{amsmath,amssymb,epsf}

\newcommand{\al}{\alpha'}
\newcommand{\de}{\partial}
\newcommand{\be}{\begin{equation}}
\newcommand{\ba}{\begin{eqnarray}}
\newcommand{\ea}{\end{eqnarray}}
\newcommand{\ee}{\end{equation}}
\newcommand{\barr}{\begin{array}}
\newcommand{\earr}{\end{array}}

\newcommand{\we}{\wedge}

\newcommand{\f}{\frac}
\newcommand{\s}{\sqrt}
\newcommand{\ti}{\tilde}
\newcommand{\ap}{\alpha}
\newcommand{\bt}{\beta}
\newcommand{\ga}{\gamma}
\newcommand{\dt}{\delta}

\newcommand{\ddd}{\cdot\cdot\cdot}
\newcommand{\no}{\nonumber \\}
\newcommand{\la}{\langle}
\newcommand{\lb}{\rangle}

\def \cA{{\cal A}}

\def \cB{{\cal B}}

\newcommand{\N}{{\mathbf{N}}}
\def \cT{{\cal T}}
\newcommand{\R}{{\mathbf{R}}}
\newcommand{\Z}{{\mathbf{Z}}}

\newcommand{\T}{{\mathbf{T}}}
\newcommand{\M}{{\mathbf{M}}}

\newcommand{\bp}{\begin{pmatrix}}
\newcommand{\ep}{\end{pmatrix}}
\newcommand{\bsp}{\begin{split}}
\newcommand{\esp}{\end{split}}

\def \cA{{\cal A}}
\def \cB{{\cal B}}
\def \cC{{\cal C}}
\def \cD{{\cal D}}

\def \rank{\mathrm{rank}}

\def \dim{\mathrm{dim}}

\def \Ed#1#2{\mbox{End}_{#1} {#2}}

\def \half{\frac{1}{2}}
\def \ov#1{\frac{1}{#1}}

\def \hF{{\hat F}}

\def \apart#1#2
{{\overset{\curvearrowright}{\partial}}_{{#1}_{#2}}^{v_{#2}({#1})
}}

\def \M{\mbox{M}}
\def \J{{\bf J}}

\begin{document}

\begin{titlepage}
\thispagestyle{empty}
\begin{flushleft}
UTMS 2001-9 \hfill hep-th/0104143 \\ UT-930
\hfill April, 2001 \\
\end{flushleft}

\vskip 1.5 cm

\begin{center}
\noindent{\Large \textbf{Exact\ Tachyon\ Condensation }}\\
\vskip 0.5cm
\noindent{\Large \textbf{on\ Noncommutative\  Torus}}
\vskip 2cm
\noindent{Hiroshige Kajiura$^a$, Yutaka Matsuo$^b$ and
Tadashi Takayanagi$^b$
}\\
{\it
\noindent{ \bigskip }\\
$^a)$ Graduate School of Mathematical Sciences, University of Tokyo \\
Komaba 3-8-1, Meguro-ku, Tokyo 153-8914, Japan\\
\noindent{\smallskip  }\\
$^b)$ Department of Physics, Faculty of Science, University of Tokyo \\
Hongo 7-3-1, Bunkyo-ku, Tokyo 113-0034, Japan\\
\noindent{ \smallskip }\\
}
\vskip 2cm

\bigskip
\end{center}
\begin{abstract}
We construct the exact noncommutative solutions on tori.
This gives an exact description of tachyon condensation
on bosonic D-branes, non-BPS D-branes and brane-antibrane systems.
We obtain various bound states of
D-branes after the tachyon condensation.
Our results show that these solutions can be generated by applying
the gauge Morita equivalence between the constant curvature
projective modules. We argue that there is  
a general framework of the noncommutative geometry
based on the notion of Morita equivalence which underlies
this specific example.
\end{abstract}
\vfill
\end{titlepage}
\vfill
\newpage

\section{Introduction}
\hspace{5mm}
D-branes have been playing the most prominent role in recent developments
of string theory. This is not only because they are 
the non-perturbative objects but also because 
they possess many interesting characteristics
which require mathematically distinguished 
descriptions. It is very interesting to try to prove that 
these various viewpoints are mutually consistent. 
Such an effort often gives profound 
relationships between physics and mathematics.

One of such examples is the quantum field theory 
on the noncommutative space arising from open string theory
in $B$ field background \cite{CoDoSc,SeWi}. 
In the commutative case, the space of functions
gives the basis of the quantum field theory,
for example, through the mode expansion.
In the noncommutative space they are replaced by
the noncommutative $C^*$-algebra  $\cA$ which is
at the heart of the principle of the noncommutative
geometry (see for example \cite{co-book}).
In this paper we are mainly  interested in
the description of D-branes with the direct use of such principle.

In the commutative approach, the soliton charges of the
D-branes are derived from the topological $K$-group
\cite{MiMo, Wi}. It is based on the observation that
there are always the massless gauge particles which define
the vector bundle on the D-branes. We need to take the 
formal difference of vector bundles
to describe the brane-antibrane pair annihilation process
\cite{Se2,Se1}.

In noncommutative geometry, the topological
information of the `manifold' is given by the operator 
algebra version of the $K$-group $K(\cA)$.
For example in Connes' index theorem \cite{co-book},
the topological index is give by the pairing of the
element of $K$-group with that of the cyclic 
cohomology group of $\cA$.
We note that $K_0(\cA)$ is defined by an equivalence class of
projection operators in $Mat_{\infty}(\cA)$, the infinite dimensional 
matrix algebra with elements in $\cA$. 

{}From physical viewpoints,
it is natural to conjecture that topological properties
of D-branes are described by the operator algebra $K$-group.
It should be described through the solitonic configurations 
which are proportional to projection operators. 
Remarkably such configurations
indeed appear as classical solutions (GMS soliton) 
in the scalar field theory on the noncommutative plane (Moyal plane) 
in the large noncommutativity limit \cite{GoMiSt}. 
This idea was immediately applied to
the string theory in the tachyon condensation process \cite{Se1,Se2},
{\it e.g.} bosonic D-branes, non-BPS D-branes and 
brane-antibrane systems \cite{HaKrLaMa,DaMuRa}.
If we use nontrivial GMS soliton,  
the lower dimensional D-branes are generated.
As pointed out in \cite{Ma,HaMo} 
it can  be seen as the noncommutative
generalization \cite{Wi4} of the correspondence 
between D-brane charges and $K$-theory  \cite{MiMo,Wi}.
 
In this way the unstable systems of D-branes 
seem to give a good example for the application of 
the geometrical methods of the noncommutative
geometry to the string theory. There are two directions to proceed.
One is to challenge noncommutative spaces with
more complicated and richer structures. The simplest nontrivial
example is the noncommutative torus.
This example is interesting from physical side since we expect
to have analogue of T-duality symmetry in the form of Morita equivalence.
It was investigated in \cite{BaKaMaTa,SaSa}
by using Powers-Rieffel projection.  Unlike the Moyal plane,
we observed a sort of instability \cite{BaKaMaTa}.
It comes from the fact that we may construct the noncommutative
soliton with arbitrary small size.  Mathematically it is related to the
fact that $K$-group of noncommutative tori is not quantized in
$\bf Z$ but takes its value in $\bf R$.  It remained as a puzzle
whether it is natural to interpret the continuous value
as the D-brane charge. Later
a different construction of the soliton configuration
on the torus and on the orbifold was discussed in \cite{MaMo}.
Among other things, a remarkable suggestion is to use Morita equivalence
bimodule in the construction of the noncommutative soliton.
Similar construction of the projection operator on tori
was also discussed in \cite{Bo} and \cite{GoHeSp}.
For fuzzy sphere, GMS-like solitons were discussed in \cite{HiNoTa}.

The other direction is to take the gauge field into account and to
construct the exact solution without taking the large $B$ limit.
It was pioneered in \cite{HaKrLa} when the base space is
Moyal plane. Certain constraint on the 
coupling of field strength and tachyon field
should be satisfied in order to have such property.

In the present paper we continue to study
the noncommutative soliton on the two-torus. 
We have mentioned two motivations, (1) how to resolve instability
of the spectrum and (2) the construction of exact solution.
The use of Morita equivalence initiated in \cite{MaMo}
gives another motivation.  For the nontrivial examples such as
tori, we can not directly construct the analogue of the shift operator.
In this sense,
we can not escape from using more 
abstract Morita equivalence bimodule directly
to construct the noncommutative soliton.  
(A nice review of Morita equivalence for noncommutative torus
is given by \cite{KoSc}). 
Once we know how to use it, one may apply the method to other
examples as well, namely
in the generic examples of the open string systems
interpolating D-branes \cite{SeWi,Wi4}.
We argue that the Morita equivalence
gives a natural generalization of the notion of the brane-antibrane
systems and leads to the description which is similar to the
superconnection \cite{Wi,KeWi,KrLa,TaTeUe}. Inspired by this fact
we propose an equation which defines the noncommutative solitons
on  brane-antibrane systems in the generalized sense. 

As we will see, we can obtain the exact solutions of 
the tachyon condensation on the two-torus
by employing the constant curvature connections.
In the noncommutative torus, the constant curvature connection
parameterizes the equivalence class of the
whole projective modules (analogue of vector bundle).
As a result we obtain various D2-D0 bound states
after the tachyon condensation on a non-BPS D2-brane.
We also find that the gauge Morita equivalence \cite{S,KoSc} 
(bimodule between constant curvature connection) plays the crucial role of 
generating solutions.  In fact one can construct exact solutions for any 
$C^*$-algebras if there exist the
gauge Morita equivalence bimodules. Furthermore this exact analysis 
explicitly shows that for finite $B/g$ the above mentioned 
instability does not occur.

The paper is organized as follows. In section 2, we discuss 
tachyon condensation on generic noncommutative spaces. After we review the
Morita equivalence, we construct the projection operator
of the brane-antibrane
systems in the generalized sense. 
We see that the structure of superconnection \cite{Qu} 
naturally appears as the linking algebra in the framework
of $C^*$-algebra.
In section 3 we discuss the tachyon condensation 
on a noncommutative two-torus. 
We construct the exact solutions for bosonic D-branes, 
non-BPS branes and brane-antibrane systems in terms of flat curvature
connection. 
We also discuss the solution generating rule for this 
examples by using the gauge Morita equivalence. 
This section also includes a review of some mathematical
results on the projective modules. 
In section 4 we summarize the conclusions. In the appendix A we give
a review of the explicit example of projections in noncommutative tori and we
also show the calculations of their topological charges.

\section{Morita equivalence and noncommutative soliton:
A  General Strategy}
\setcounter{equation}{0}

We start from discussing relatively formal viewpoint which
will be useful in the later sections.  While our main
result is restricted to the noncommutative solitons on
noncommutative tori in section 3, 
we think that our method can be basically
applied to the other open string systems such as $Dp-Dp'$ 
as well.

Let us first recall the definition of the noncommutative solitons 
\cite{GoMiSt}.
They are the solutions to the equation of motion,
\begin{equation}
 \frac{\partial V(\star \phi)}{\partial \phi}=0.
\end{equation}
Here $\star$ is the (noncommutative) product of
the given  $C^*$-algebra $\cA^{(0)}$ which defines the noncommutative
geometry on the single D-brane.
It is solved in the following form,
\begin{equation}
 \phi(x)=\sum_{i} \lambda_i P_i,
\quad
\lambda_i \in {\bf R},
\quad
P_i\in \cA^{(0)}\ ,
\end{equation}
where $\lambda_i$s are the solutions to the equation
$\partial V(\lambda)/\partial \lambda=0$ and $P_i$'s are
the mutually orthogonal projections
$P_i \cdot P_j=\delta_{ij}P_i$, $P_i^*=P_i$. 
In this sense, the construction of noncommutative solitons
is reduced to find the projection operators
\footnote{If the number of the D-branes is greater
than one (say $N$), we need to consider the projector
in the matrix algebra  $Mat_N(\cA)$.}. 

In the mathematical context, the classification 
of the projection operator is directly related to the definition
of the $K_0$-group of the operator algebra $K$-theory. 
This is related to the fact that the noncommutative analogue
of the vector bundle is described by the projective module of
given $C^*$-algebra $\cA$.
Let us briefly illustrate the correspondence
by using the relationship with the commutative 
theory. 

Let $M$ be a smooth manifold (base space) 
and $E$ be a vector bundle over $M$. 
In the context of the topological $K$-theory, 
it is known that the isomorphism class of $E$ is an element 
of $K^0$-group\footnote{
In general, the formal difference of the isomorphism class 
of vector bundles over $M$ is the element of $K^0$. }. 
By Swan's theorem, for any $E$ there exists vector bundle $F$ such that 
$E\oplus F$ is a trivial bundle over $M$. 
Let $C^\infty(M)$ be the smooth function on $M$.
The trivial bundle can be written as $(C^\infty(M))^N$ with some
$N\in\N$. Therefore, any vector bundle over $M$ can be obtained by
acting $P$ on $(C^\infty(M))^N$, where $P$ is a projection 
in $Mat_N(C^\infty(M))$\footnote{
We define the algebra $Mat_N(\cA)$ as the $N$ times $N$ matrix algebra
with elements in $\cA$.}. The module  which is constructed by
applying the projection operator to the free module is called
projective module.

This characterization of the vector bundle
can be generalized to noncommutative theory. 
A noncommutative algebra $\cA^{(0)}$ replaces $C^\infty(M)$
for `noncommutative space'. 
The free module $(\cA^{(0)})^N$ 
corresponds to the $\rank\ N$ trivial bundle on commutative space. 
Projective module $E$ is defined as $\cA^{(0)}$-module such that 
there exists the other $\cA^{(0)}$-module $F$ with 
$E\oplus F=(\cA^{(0)})^N$. 
Thus the noncommutative analogue of a vector bundle is 
the projective module $E=P(\cA^{(0)})^N$ which is defined 
by a projection operator $P\in Mat_N(\cA^{(0)})$.
In this sense, D-branes on noncommutative spaces are described by 
the projective modules, and the operator algebra $K_0$-group
classifies the D-branes on the noncommutative space \cite{Wi4,Ma,HaMo}.

Let us come back to the issue of the construction of the projection
operator.  We would like to use the Morita equivalence bimodule
as the abstract building block to construct noncommutative soliton.

Morita equivalence is one of the central idea of the classification
of $C^*$-algebra.
{}From the  mathematical viewpoint, it is essential to determine
when two $C^*$-algebras ${\cal A, B}$ define the 
same type of noncommutative geometry.  In the noncommutative
geometry, the idea of points is replaced by the set of ideals
of $C^*$-algebra.
It is then known that two $C^*$-algebras ${\cal A, B}$
have the same set of ideals if there are ${\cal A}$-${\cal B}$ Morita
equivalence bimodule ${}_{\cal A}X_{\cal B}$ 
(for example see \cite{morita}). It is defined
as a bimodule on which ${\cal A}$ (resp. ${\cal B}$) acts 
from the left (resp. right) with 
two types of inner products $\langle\ ,\ \rangle_{\cal A}$,
$\langle\ ,\ \rangle_{\cal B}$ of  ${}_{\cal A}X_{\cal B}$
with value in $\cal A$ and $\cal B$, respectively
with the following conditions
\begin{eqnarray}
 \langle a x, y\rangle_{\cal A}=a\langle x, y\rangle_{\cal A}\ ,\quad
 &\langle x, y\rangle_{\cal A}^*=\langle y, x\rangle_{\cal A}&\ ,
 \qquad a\in {\cal A}\ ,\nonumber\\
 \langle x, yb\rangle_{\cal B}=\langle x, y\rangle_{\cal B}b\ ,\quad
 &\langle x, y\rangle_{\cal B}^*=\langle y, x\rangle_{\cal B}&\ ,
 \qquad b\in {\cal B} 
 \ ,\label{inner1}
\end{eqnarray}
\begin{eqnarray}
 \langle xb, y\rangle_\cA=\langle x, yb^*\rangle_\cA\ ,\qquad
 \langle ax, y\rangle_\cB=\langle x,a^* y\rangle_\cB\ .\label{inner2}
\end{eqnarray}
The most important property
which should be satisfied by them is the 
associativity  
\begin{equation}
 \langle x,y\rangle_{\cal A} z = x \langle y, z\rangle_\cB,
\quad x, y, z\in {}_{\cal A}X_\cB.\label{bimodule2}
\end{equation}
Two $C^*$-algebras $\cA$ and $\cB$ 
which have such a Morita equivalence bimodule ${}_{\cal A}X_{\cal B}$ 
is called Morita equivalent.

In the string theory, there is a natural interpretation
of such equivalence relation. 
It is well-known that the
noncommutativity arises in the string theory on the
D-branes connected by the open string in the presence of
$B$ field. On the two ends of open string, 
we have two D-branes and generally two different types
of noncommutative geometry defined on them.
Suppose they are defined by the $C^*$-algebras $\cal A, B$
\footnote{More precisely, 
let $\cA^{(0)}$ be the algebra corresponding to the noncommutative 
base space, and 
let $E_\alpha$ (resp. $E_\beta$) be the projective $\cA^{(0)}$-module (D-brane)
related to $\cA$ (resp. $\cB$), then $\cA=\mbox{End}_{\cA^{(0)}}{E_\alpha}$ 
and $\cB=\mbox{End}_{\cA^{(0)}}{E_\beta}$.
Let $a\in\cA^{(0)}$, $A\in\Ed{\cA^{(0)}}{E}$ and $\xi\in E$.
$\Ed{\cA^{(0)}}{E}$ means $(a\cdot\xi)A=a(\xi\cdot A)$. 
This is 
the natural noncommutative generalization 
of the definition of endomorphisms for vector bundles.}. 
Under such circumstances, it is natural to conjecture
(for example, see \cite{SeWi,Wi4}),
\begin{itemize}
 \item The bimodule naturally interpreted as the
open string field $\Psi$ where $\cA$ (resp. $\cB$)  
acts from the left (resp. right).  
\item Two inner products are identified the product of
open string fields. We have two type of inner product because
we have two choices (which side of the open string) for the contraction.
\item The associativity of the product corresponds 
to that of the product of the open strings.
\end{itemize}
Although the actual justification of these statements is far from
being obvious at this stage, it gives a nice intuition to
the otherwise abstract nature of Morita equivalence.

In the following, we use
the Morita equivalence bimodule to define the
projection operator (= noncommutative soliton).
In an abstract language, it can be described as 
follows \cite{Ri1,Ri2,morita,MaMo}.
In the very definition of the Morita equivalence, we actually
need to impose that the inner product $\langle\ ,\ \rangle_{\cal A, B}$ 
maps ${}_{\cal A}X_{\cal B}$ into a dense set of the $\cal A, B$
respectively.  Namely for any operator $a\in\cal A$, there should be
a finite set $x_i, y_i\in {}_{\cal A}X_{\cal B}$ such that 
$a=\sum_i\langle x_i, y_i \rangle_{\cal A}$.
Suppose $\cal A$ has identity as its element
and take $1=\sum_{i=1}^N\langle x_i, y_i \rangle_{\cal A}$.
{}From $x_i$, $y_i$ one may define the projection operator
in $Mat_N(\cB)$  as 
$P\equiv\langle y_i, x_j \rangle_{\cal B}$
since
\begin{equation}
 \sum_j\langle y_i, x_j \rangle_{\cal B}\langle y_j, x_k \rangle_{\cal B}
 =\sum_j\langle y_i, x_j \langle y_j, x_k \rangle_{\cal B}\rangle_{\cal B}
 =\langle y_i, \sum_j\langle x_j, y_j\rangle_{\cal A} x_k \rangle_{\cal B}
 =\langle y_i, x_k \rangle_{\cal B}.
\end{equation}
Unlike the original GMS soliton where the projection operator
defines the lower dimensional D-branes, it seems rather hard
to identify the nature of the projected space.  However, from 
the mathematical side,
$\cal A$ can be embedded into $Mat_N(\cal B)$, as 
\begin{equation}
 {\cal A} \sim P \cdot Mat_N({\cal B})\cdot P\,\,.
\end{equation}
Suppose $\cal A$ (resp. $\cal B$) describes the noncommutative geometry 
on D-brane $\alpha$ (resp. $\beta$). Then
a physical interpretation of above identity is that $\alpha$
can appear as the noncommutative soliton on the
$N$ copies of the D-branes $\beta$ through the tachyon condensation
process in a generalized sense.

In order that this type of interpretation is possible, we need the analogue
of $D\bar{D}$ pair for this generalized setting.
In the study of $D\bar{D}$ pair, the tachyon condensation process
is described by the combination of gauge fields and tachyon fields
(so called `superconnection' \cite{Qu}) as argued in 
\cite{Wi,KeWi,KrLa,TaTeUe},
\begin{equation}\label{superconnection}
 \left(
 \begin{array}{cc}
  d+A_1 & T\\ \bar{T} & d+A_2
 \end{array}
 \right)
\end{equation}
where each entry represents the various sectors of the open string.

There exists an analogue of superconnection in $C^*$-algebra
which is called the linking algebra $\cal C$
(see for example, \cite{morita}).
It is defined in such way as containing the algebras $\cal A, B$
as its complementary components.  Namely there exists a
projection $P\in \cal C$ such that ${\cal A}=P\cdot {\cal C}\cdot P$
and ${\cal B}=(1-P)\cdot {\cal C} \cdot (1-P)$.
It is known that the linking algebra $\cal C$ exists
if and only if two $C^*$-algebras ${\cal A, B}$ are Morita equivalent.
Namely if the bimodule ${}_{\cal A}X_{\cal B}$ exists, one may define
the linking algebra by two by two matrices,
\begin{equation}\label{linking-algebra}
 \left(
 \begin{array}{cc}
  a & x\\ \bar{y} & b
 \end{array}
 \right)
 \quad
 a\in {\cal A},\,\,
 b\in {\cal B},\,\,
 x, y\in {}_{\cal A}X_{\cal B}.
\end{equation}
One may easily check that the matrix multiplication 
of such $2\times 2$ matrices is well-defined
and the obvious projection operator
$P=\left(\begin{array}{cc}1 & 0 \\ 0 & 0 \end{array}\right)$
gives the projection into $\cal A$.
By comparing (\ref{superconnection}) and (\ref{linking-algebra})
one sees that the Morita equivalence bimodule plays the analogous r\^ole
as that of the tachyon field and that the two gauge fields correspond to
 the endomorphism algebras $\cA=\Ed{\cA^{(0)}}{E_\alpha}$ and 
$\cB=\Ed{\cA^{(0)}}{E_\beta}$.

This analogy was implicitly used in the explicit calculation
of the tachyon condensation of $D\bar{D}$ system in the Moyal plane
\cite{Wi3,HaKrLa}. We summarize our conventions for Moyal plane in 
the appendix A.
We recall that the noncommutative soliton in this case was constructed
out of the partial isometry described by the tachyon
fields $T$ and $\bar T$ which satisfy,
\begin{equation}\label{partial}
 T \bar T T = T, \qquad
 \bar T  T \bar T = \bar T.
\end{equation}
With these relations, $T \bar T$ and $\bar T T$ becomes projection operators
and defines the noncommutative soliton.  Explicit form of $T$ can be constructed
by using the shift operator $S=\sum_n |n\rangle \langle n+1|$,
which satisfies $S \bar S =1$ and $\bar S S=1-|0\rangle \langle 0|$ and 
so on \cite{HaKrLa}. 
The tachyon field is given by $T=S^m \ (m\in\N)$ with this solution
generating rule. 
In this case, $T$ gives the isometry between the subsets of the
identical algebra $\cA$ and $\cA$ acts on $T$ from the both side.

In our more general situation, the r\^ole of $T$ resembles 
that of the Moyal plane case
but the algebra acting on $T$ from left is in general different from
that acting from right. Thus $T$ should be 
an element of an equivalence bimodule.
If we replace the partial isometry by the Morita equivalence bimodule,
the statement which corresponds to (\ref{partial}) is 
that one may
choose an element $T$ of the bimodule which satisfies
\begin{enumerate}
 \item $P=\langle T, T\rangle_{\cal A}$ and $Q=\langle T, T\rangle_{\cal B}$
 are the projectors of $\cal A, B$.
 \item It satisfies the analog of the partial isometry relation,
\begin{equation}
 \langle T, T\rangle_{\cal A}T = T \langle T, T\rangle_{\cal B}= T\ .
 \label{piso}
\end{equation}
\end{enumerate}
While it is a nontrivial question whether we may find such $T$,
the requirement of these two conditions are actually 
equivalent \cite{Ri1}. 
Suppose that the partial isometry-like equation (\ref{piso}) holds.
$P=\langle T, T\rangle_\cA$ and $Q=\langle T, T\rangle_\cB$ are clearly 
self-adjoint, and direct calculation shows that $P^2=P$ and $Q^2=Q$ 
by using the properties of the two inner products
(\ref{inner1})(\ref{inner2})(\ref{bimodule2}). 
Conversely, 
if we start from
the condition $P^2=P$, the partial isometry-like equations follows
from vanishing of the norm,
\begin{equation}
 \langle \langle T, T\rangle_{\cal A}T - T,
 \langle T, T\rangle_{\cal A}T - T\rangle_{\cal A}=0.
\end{equation}
One can also see from this proof that $P^2=P$ implies $Q^2=Q$.

We would like to propose that the equation (\ref{piso}) defines the 
noncommutative soliton on the brane-antibrane systems
in the generalized sense. We will see
the explicit examples on noncommutative tori in section 3.4.

\section{Noncommutative Torus and D-branes}
\hspace{5mm}
Because our discussions so far are  given in the abstract language, 
it is desirable to investigate
explicit examples in order to illuminate the idea. The simplest
example is D-branes on noncommutative plane (Moyal plane)
but it is too simple to use the machinery we would like to examine
since it reduces to the (infinite dimensional) matrix algebra.
Thus in this section we consider D-branes 
on noncommutative tori, where Morita equivalence is interpreted as
T-duality   \cite{CoDoSc,S,RS,PS} and has a rich structure.

The algebra of noncommutative two-torus is 
generated by unitary elements $U_1$ and $U_2$ with the relation,
\ba
U_1U_2=U_2U_1e^{2\pi i\theta},
\ea
where the real number $\theta\in [0,1]$ is the parameter of the algebra.
We write this algebra by $\cA_{\theta}$. The generators can be written
in terms of noncommutative coordinates 
$(x^1,x^2)$ with $[x^1,x^2]=-2\pi i\al\theta$
as follows
\ba
U_1=e^{ix^1/\s{\al}},\ \ \ U_2=e^{ix^2/\s{\al}}.
\ea
A generic element $a\in \cA_{\theta }$ can be expanded 
by $U_i$  as\footnote{%
To be exact we should assume $a_{mn}$ tend to zero faster than any 
powers of $\left|
n\right| +\left| m\right| $ as $\left| n\right| +\left| m\right| \rightarrow
\infty .$}
\begin{equation}
a=\sum_{m,n\in {\Z}}a_{mn}U_1^{m}U_2^{n}\ .
\end{equation}
While we can not realize this algebra as the matrix algebra
for the irrational $\theta$, one may
formally define the trace for $\cA_{\theta}$ as follows
\begin{equation}
\mbox{Tr}\,\,a=a_{00},
\end{equation}
by using the above expansion. This is equal to the integration
over $(x^1,x^2)$. It is obvious that it satisfies the fundamental
relation $\mbox{Tr}(ab)=\mbox{Tr}(ba)$.

As we saw in section 2, the gauge bundle on the D-brane
is described by the projective $\cA_\theta$-module.
It is known that the isomorphic class of the projective 
modules on a noncommutative torus 
is classified by their Chern characters \cite{Ri2}
which specifies the element of $K_0(\cA_{\theta})$.
The relation of these mathematical facts to the  
RR-couplings of brane-antibrane systems will be discussed later
in the subsection 3.4.

In this section, after explaining some mathematical backgrounds on 
the projective module on the noncommutative tori, 
we discuss mass spectrum of BPS D-branes
and their T-duality transformation. Finally we investigate tachyon 
condensation on non-BPS D-branes and brane-antibrane systems.

\subsection{Projective modules, Constant Curvature Connection 
and Morita equivalence
on Noncommutative Torus}
\hspace{5mm}
In this subsection, we explain some mathematical results
on $\cA_\theta$, especially the projections, the projective modules, 
the constant curvature connections and 
Morita equivalence%
\footnote{See \cite{KoSc} for an more pedagogical review.}
which will be essential for later arguments. 
The readers who is familiar with the subject may
skip this subsection.

We start from defining the projection operators on $\cA_{\theta}$.
In this algebra the unitary equivalence class\footnote{
A projection $P$ is said to be unitary equivalent to another projection $Q$
 if an unitary element $u\in \cA$ exists such that $P=u^*Qu$.} of projections 
is characterized by the value of the trace $\mbox{Tr}$ 
\cite{Ri1,PV2} as follows
\ba
\mbox{Tr}(P_{n-m\theta})=n-m\theta,\ \ \ \ (0\leq n-m\theta\leq 1),
\label{prj}
\ea
where for each pair of integers $n$,$m$ which satisfy $0\leq n-m\theta\leq 1$ 
there exists a equivalence class of projections and we wrote an element 
in this class as $P_{n-m\theta}$. In other words $K_{0}$-group 
$K_{0}(\cA_{\theta})$ is given by
\ba
K_{0}(\cA_{\theta})={\bf Z}+\theta{\bf Z}\in {\bf R}\ .  \label{k-th}
\ea
The explicit construction of projections was given in \cite{Ri1} and is
called Powers-Rieffel projection. Some details of this construction
are reviewed in the appendix A.

Let us turn to projective modules on the noncommutative torus $\cA_{\theta}$. 
%
Among various projective modules, we are interested 
in those which has the constant curvature connection.
The explicit construction can be found in \cite{Co,Ri2,CoDoSc,KoSc} and
we review this below.


Starting from a set of the
rapidly decreasing functions on $\bf R$, $\phi_j(x)$,
$j\in {\bf Z}_m$. We define the right action of $U_i$ ($i=1,2$)
as follows,
\begin{eqnarray}
 ( \phi U_1)_j(x)& = & \phi_{j-1}(x+\frac{n}{m}-\theta)\nonumber\\
 (\phi U_2)_j(x) & = & \phi_{j}(x)e^{2\pi i(x+jn/m)}.
\end{eqnarray}
One may easily confirm that the operators $U_1, U_2$ thus defined
satisfies $U_1 U_2=e^{2\pi i \theta}U_2 U_1$. 
It shows that the Schwartz space is right $\cA_\theta$ module.
It contains two integer parameters $n\in {\bf Z}, m\in {\bf Z}\geq 0$ 
and we denote it as
$E_{n,m}$.
It has the following properties,
\begin{enumerate}
 \item The natural `covariant derivative' for this module is given by
 \begin{equation}
  \nabla_1=-\frac{2\pi i m}{n-m\theta} x,
  \quad
  \nabla_2=\frac{\partial}{\partial x},
 \end{equation}
 which satisfies $\nabla_i U_j=2\pi i\delta_{ij}U_j$.
 It has the constant curvature,
 \begin{equation}
  F_{12}=-F_{21}=\frac{2\pi i m}{n-m\theta}. \label{constc}
  \end{equation}
  \item Its endomorphism $End_{\cA_{\theta}}E_{n,m}$ 
  is given by $\cA_{\ti \theta}$ ($\ti{\theta}
  = -\frac{b-a\theta}{n-m\theta}$). This acts on $E_{n,m}$ from the left. 
  The integers $a,b$ are determined
  from $n,m$ by the condition $an-bm=1$.
  The generators of $\cA_{\ti\theta}$
  act on them as
  \begin{eqnarray}
   (Z_1\phi )_j(x) & = & \phi_{j-a}(x+1/m)\nonumber\\
   (Z_2\phi )_j(x) & = & \phi_{j}(x)e^{2\pi i(
\frac{x}{n-m\theta}+\frac{j}{m})}\ .
   \label{thetat}
  \end{eqnarray}
  One may easily confirm that the action of $Z_i$ are compatible
  with the action of $U_i$ since $[Z_i, U_j]=0$. 
   Note that if the module is free $(n,m)=(1,0)$,
  then $\ti{\theta}=\theta$.
 \end{enumerate}
Therefore the Schwartz space defines the Morita equivalence bimodule
between $\cA_\theta$ and $\cA_{\ti\theta}$. More generally, if the integers
 $(n,m)$ is not coprime, then one can construct the reducible
  projective module $E_{n,m}=E_{n/d,m/d}\oplus E_{n/d,m/d}\oplus \ddd
  \oplus E_{n/d,m/d}$, where we have defined $d=\mbox{g.c.d}(n,m)$. From this
 one can conclude that $\cA_{\ti\theta}$ is Morita equivalent to
 $Mat_{d}(\cA_{\theta})$ if there exist coprime integers
$(n_0,m_0)\equiv(n/d,m/d)$ and $(a,b)$ 
 such that
 $\ti{\theta}=\frac{b-a\theta}{n_0-m_0\theta}$ 
and $an_0-bm_0=-1$ \cite{Ri1}. 

It will be useful
to define the explicit form of $\cA=\cA_{\ti\theta}$ and $\cB=\cA_\theta$
inner product for the Schwartz space. This problem is solved in
much more general sense by Rieffel \cite{Ri2}.  
However we write down some of the
explicit forms for our specific
example
\footnote{General formula for the generic Heisenberg module
is not difficult to write down.  One outline is
sketched in Appendix A of \cite{MaMo}.}. 
For $\phi, \psi\in S({\bf R}\times {\bf Z}_m)$,
we define,
\begin{eqnarray}
 \langle \phi, \psi \rangle_\cA & = & 
 \frac{1}{n-m\theta}\sum_{m,n}(Z_2^{-n}Z_1^{-m}\phi, \psi)_\cA
 Z_1^m Z_2^n\\
 \langle \phi, \psi \rangle_\cB & = & 
 \sum_{m,n}(\phi,\psi U_2^{-n}U_1^{-m})_{\cB}
 U_1^m U_2^n\nonumber\\
 (\phi,\psi)_{\cB} & = & (\psi,\phi)_{\cA} = \int_{-\infty}^\infty dx
 \sum_{i\in {\bf Z}_m} \overline{\phi_i(x)}\psi_i(x).
\end{eqnarray}
The basic properties of inner product (\ref{inner1}) and
 (\ref{inner2})
 follows from the
identities $(a\phi ,\psi)_{\cA,\cB}=(\phi,a^{*}\psi )_{\cA,\cB}$, 
$(\phi \,b,\psi)_{\cA,\cB}=(\phi,\psi\,b^*)_{\cA,\cB}$ and so on. 
The derivation 
of the associativity (\ref{bimodule2}) is much more nontrivial.  
We need to use
the Poisson resummation formula,
\begin{equation}
 \sum_{n=-\infty}^\infty f(\alpha n)=
 \frac{1}{\alpha}\sum_{m=-\infty}^\infty \tilde{f}\left(
\frac{2\pi m}{\alpha}\right)\qquad
 \tilde{f}(m)= \int_{-\infty}^\infty e^{imx}f(x)dx.
\end{equation}
After some calculations, one may confirm,
\begin{equation}
 (\langle \phi,\psi\rangle_{\cA}\chi)_i(x) = 
 (\phi\langle \psi,\chi\rangle_\cB)_i(x)=
 \sum_{r,s\in{\bf Z}} (\phi U_1)^r_i(x) \overline{
 (Z_1^s \psi U_1^r)}_i(x) (Z_1^s\chi)_i(x).
\end{equation}

Among all connections on a given projective module,
the constant curvature connection is the most useful in the physical
application. One reason for this is that such a special 
connection appears as the 
solution of the BPS equation \cite{S,KoSc}. Another reason will be
given later in the arguments of the exact solution
for the tachyon condensation. 


Finally we would like to add a comment on the 
topological invariants. From the constant curvature
connection, it is rather easy to evaluate the
Chern character  \cite{Ri2,S} and it is known that
each component becomes integer after the modification
similar to Myers term \cite{My}\footnote{
This modification can also be regarded as `quantum effect' 
in terms of the quantum twisted bundle \cite{Twist}. 
This procedure shows the relation between a bundle on 
the commutative torus and that on the noncommutative torus geometrically, 
and agrees with so-called Seiberg-Witten map\cite{SeWi}. 
We will use this map in eq.(\ref{swm}).}.
It is of some interest to confirm this fact by using other
form of the connection/curvature.
In general a connection on projective module
(``Levi-Civita connection'')
is constructed in the form (see for example \cite{Co, KoSc,Ma})
$\nabla_i=P_{n-m\theta}\cdot\delta_i\cdot P_{n-m\theta}$.
The trace of the curvature (first Chern class) is the cyclic 
2-cocycle $\tau_2$ for $P_{n-m\theta}$. 
In appendix A we evaluate it by using the Powers-Rieffel
projection as $P_{n-m\theta}$ and derive $\tau_2=m$.
This is of course consistent with the computation from
the constant curvature connection  (\ref{constc}).





\subsection{Noncommutative Description of BPS D-branes and T-duality}
\hspace{5mm}
Here we discuss the spectrum 
and the T-duality transformation rule of the BPS D-branes 
on a two dimensional torus with a $B$-field flux. 
As is well-known, there are two viewpoints for this system. 
One is the conventional description (commutative description) 
using the closed string variables. 
The other is by the open string variables with
the noncommutative geometry \cite{CoDoSc,SeWi}. 
The results given here will be useful in the later discussions. 
Some general arguments can be found in \cite{HofVer1,KoSc}.

We investigate the dynamics of D-branes on the
noncommutative two-torus $\cA_{\theta}$.
Since we restrict our interest to the two dimensional case,
we discuss D2-D0 bound states on the torus below.
If the D2 charge and D0 charge are given by $(n,-m)$, the mass of
D2-D0 bound state is determined as follows
\ba
 \M^{BPS}_{(n,m)}=\f{|n|}{ \s{\al} g_s}\s{\det\left(g+2\pi\al(B+F)\right)},
 \label{bpsnm}
\ea
where we have defined the gauge field strength
\ba
 F=\f{\J}{2\pi\al}\f{m}{n}\ ,\qquad
 \J\equiv\bp 0&-1\\ 1 & 0\ep\ \label{flux}.
\ea

In this formula we have taken the effect of $-m$ D0-branes into 
account as the shift of gauge field strength. 
Note that the value of $n$ or $-m$ can be negative integer
because of T-duality. 
For a review of T-duality on general tori see \cite{GPR}.
In the following we examine the transformations from a 
D2-brane to various D2-D0 bound states.

We define $E=g+2\pi\alpha'B \in Mat_2(\R)$. A single 
D$2$-brane has the mass,
\ba
 \M^{BPS}_{(1,0)}=\ov{g_s\sqrt{\alpha'}}\sqrt{\det(E)}\ .
\ea
The T-duality group on $\T^2$ is given by $SO(2,2 |\Z)$ and this acts on 
$g, B$ and $g_s$ as follows \cite{GPR}
\ba
 \ti{E}&=&\cT(E)=\left(\cA E+\cB\right)\left(\cC E+\cD\right)^{-1}\\
 \ti{g_s}&=&\cT(g_s)=g_s\det{(\cC E+\cD)^{-\half}}\\
  \cT&=&\bp \cA &\cB \\ \cC & \cD \ep \in SO(2,2 |\Z).
 \label{T}
\ea
The mass $\M^{BPS}_{(1,0)}$ transforms into the following form
\ba
 \cT(\M^{BPS}_{(1,0)})&
 ={\displaystyle \ov{\cT(g_s)\sqrt{\alpha'}}\sqrt{\det \cT(E)}}\no
 &={\displaystyle \ov{g_s\sqrt{\alpha'}}\sqrt{\det (\cA E+\cB)}}\ .
 \label{D}
\ea
When the dimension of the torus is two, the T-duality 
group can be decomposed as $SO(2,2 |\Z)\simeq SL(2,\Z)\times SL(2,\Z)$. 
One of two $SL(2,\Z)$ groups is the 
modular transformation of the target space $\T^2$. Because it
preserves $g_s$ and $\sqrt{\det(E)}$, we will not consider this part. 
The other $SL(2,\Z)$ is related to Morita equivalence for noncommutative 
$\T^2$ \cite{CoDoSc,S,RS,PS} and we concentrate on this part. 
This $SL(2,\Z)$ transformation can be embedded into 
$SO(2,2 |\Z)$ as follows
\ba
 \barr{ccc}
  SL(2,\Z) & \hookrightarrow & SO(2,2 |\Z)\\
  \bp n&m\\-b&-a \ep &\mapsto  & \bp n{\bf 1}&m \J\\ b\J& -a{\bf 1}\ep .
 \earr\ \label{I}
\ea
Applying this transformation eq.(\ref{D}) can be rewritten as 
\ba
 \cT(\M^{BPS}_{(1,0)})
 =\ov{g_s\sqrt{\alpha'}}\sqrt{\det(n(g+2\pi\alpha'B)+m\J)}\ .
\ea
This is the same value as (\ref{bpsnm}) and confirms that the 
D$2$-brane mass in the background $\ti{E}=T(E)$ is equal to 
the mass of $(n,-m)$ D2-D0 bound state. Note also that the mass for
$(n,-m)$ is equal to that for $(-n,m)$. This implies that 
these two configurations should be 
an identical state and we can 
restrict the integers $(n,-m)$ to $n-m\theta \geq 0$. 

We translate these results into the noncommutative description. 
For simplicity, we fix the choice of the parameter $\Phi$ 
\cite{S,PS,SeWi} as $B=-\Phi$. The relation between the variables 
in the open and closed string theories is given \cite{SeWi} as follows
\ba
B=-\Phi=-\f{1}{2\pi\al\theta}\J,\ \ \ G=-(2\pi\al)^2B\f{1}{g}B,\ \ \ 
G_{s}=g_s \det(2\pi\al Bg^{-1})^{\f12}. \label{o-c}
\ea
This map is defined so that the mass of the single D$2$-brane coincides 
in both (open/ closed) descriptions. 
The transformed backgrounds $G$ and $G_{s}$ are called open string metric 
and open string coupling respectively.

When the field strength $F$ is constant, the field strength $\hat{F}$ 
in the open string description\footnote{
Note that the curvature in mathematical conventions in section 3.1
is related to the field strength $\hat{F}$ here
such that $F_{ij}=-4\pi ^2 i\al\hat{F}_{ij}$.} 
is given by \cite{SeWi} 
\ba
\hat{F}=\f{F}{1+2\pi\al\theta\J F}. \label{swm}
\ea
If we apply this to the D0-D2 bound states, the flux (\ref{flux}) 
is transformed into 
\ba
\hat{F}=\ov{2\pi\al}\f{m}{n-m\theta}\J.
\ea
This is the same as the constant curvature connection on the projective 
module $E_{n,m}$ reviewed
in the previous subsection.

Furthermore, the mass of D0-D2 bound states can also be
written in terms of the open string variables as follows
\ba
\cT(\M^{BPS}_{1,0})=\M^{BPS}_{(n,m)}&=&\f{|n|}{ \s{\al} g_s}
 \s{\det\left(g+2\pi\al(B+F)\right)}\no
 &=&\f{|n|}{\s{\al} g_s}\s{\det\left(g-(\f{1}{\theta}-\f{m}{n})\J\right)}
 \no
 &=&\f{\det(2\pi\al Bg^{-1})^{-\f12}}{\s{\al} g_s}
 \s{\det\left(-(2\pi\al)^2
 Bg^{-1}B(n-\theta m)+\f{n}{\theta}\J\right)} \no
 &=&\f{n-m\theta}{\s{\al} G_s}
 \s{\det(G+2\pi\al(F+\Phi))}. \label{BPSnm}
\ea
Note that the rank $|n|$ of the gauge field in the 
commutative description is replaced 
with the non-integer `rank' $n-m\theta\geq 0$. 
In noncommutative geometry such an 
appearance of non-integers is not surprising but very natural.
Indeed this is equal to the 
dimension of the projective module $\dim(E_{n,m})=n-m\theta$. 
As we will 
show later, it appears naturally in the processes of 
tachyon condensation. We also comment that the above formula 
is correct for any choice of $\Phi$.

Finally we present an interpretation of 
the above results from the viewpoints of 
T-duality $\cT$ on the noncommutative side.
It is derived by rewriting the action $\cT$ in 
(\ref{I}) in terms of the open string 
variables in (\ref{o-c})
\cite{S,PS,KoSc},
\ba
&&\ti{\theta}=\f{b-a\theta}{n-m\theta}, \ \ 
\ti{G}_{\mu\nu}=(n-m\theta)^2G_{\mu\nu},\no
&&\ti{G}_{s}=(n-m\theta)G_{s}, \ \ 2\pi\al\ti{\Phi}=(n-m\theta)^2
(2\pi\al\Phi +\frac{m}{n-m\theta}\J).\label{openT}
\ea
The transformation for $\theta$ is exactly the same 
as the Morita equivalence (\ref{thetat}).

By the definition of the transformation (\ref{openT}), 
the map from the closed string variable to the
open string variable (\ref{o-c}) and the action of the T-duality 
group on both (open /closed ) sides (\ref{T}) and (\ref{openT}) 
are compatible.
Therefore the mass of a bound state
$\M^{BPS}_{(n,m)}$ in the last line of (\ref{BPSnm}) 
can also be obtained by acting the transformation 
(\ref{openT}) on $\M^{BPS}_{(1,0)}$
in the open string variables
\ba
 \M^{BPS}_{(n,m)}
 =\ov{\s{\al} \ti{G_s}}\s{\det(\ti{G}+2\pi\al\ti{\Phi})}
 =\f{n-m\theta}{\s{\al} G_s}\s{\det(G+2\pi\al(\hF+\Phi))}\ .
\ea
This shows that a $(n,-m)$ D-brane on the noncommutative torus
$\cA_{\theta}$ is a single D$2$-brane on $\cA_{\ti\theta}$. 
This result is consistent with the previous arguments in the commutative
(closed string) side. 
Notice that the curvature $\hF$ on the noncommutative torus $\cA_\theta$ 
vanishes on the corresponding single D$2$-brane on $\cA_{\ti\theta}$ due to 
the shift of $\Phi$ in eq.(\ref{openT}).

The above arguments of T-duality can also be applied to non-BPS D-branes 
and brane-antibrane systems in the same way.  
We will see later that 
this T-duality on the noncommutative side is
more directly related to Morita equivalence in the arguments of tachyon 
condensation.

\subsection{Tachyon condensation on non-BPS D-branes}
\hspace{5mm}
Let us discuss the tachyon condensation on non-BPS D-branes (see for 
example \cite{Se2}) on the noncommutative torus $\cA_{\theta}$. 
The same arguments can be applied to the bosonic string.
Because any D2-D0 bound state of non-BPS D-branes can be 
transformed into a D2-brane, 
we can begin  with a non-BPS D2-brane. 
The relation between the variables in open 
and closed string theories is the same as (\ref{o-c}) 
and we continue to choose the value of $\Phi$ as $\Phi=-B$
to obtain the simplest expression.
The solutions, however, do hold without any modification
for general values of $\Phi$ with somewhat lengthy calculations.

On any non-BPS D-brane there exists\footnote{
In this paper we discuss the cases where
 the transverse scalars do not have expectation values.} 
a (real scalar) tachyon field $T$ and 
a gauge field $A_\mu$. As argued 
 in \cite{Se3} the effective action of
 a non-BPS D2-brane can be written as
\ba
S=\f{\s{2}}{\s{\al} G_s}\int dt\mbox{Tr}\left[V(T)
\s{\det(G+2\pi\al(\hat{F}+\Phi))}
\right]+O([\nabla,T],\ [\nabla,\hat{F}]), \label{sn}
\ea
where $[\nabla,T],\ [\nabla,\hat{F}]$ denote the covariant 
derivative of the tachyon field $T$ and the gauge
field strength $F$; the symbol $O([\nabla,T],[\nabla,\hat{F}])$ 
means those terms 
which include one or more derivatives of $T$ and $\hat{F}$. 
As we will see below our exact arguments of
tachyon condensation do not depend on the detailed form of 
the derivative terms. The factor $V(T)$ in front of the Born-Infeld term
represents the tachyon potential. We normalized the value of
the tachyon potential such that
its value before and after the tachyon
condensation into the vacuum are given by $V(1)=1$ and $V(0)=0$
following from Sen' conjecture \cite{Se1,Se2}. 
  
We use here the open string variable and therefore all the fields 
on the brane are regarded as the operators on the noncommutative torus 
$\cA_{\theta}$.
On the non-BPS D-brane, the tachyon and the gauge field
belongs to the adjoint representation of the gauge group.
In the language of the noncommutative geometry 
they are expressed as elements in the endomorphism 
$\mbox{End}_{\cA^{(0)}}E$ of the projective module $E$.
Before the tachyon condensation, the projective module should 
represent the original D2-brane itself
$E_{1,0}\ =\cA_{\theta}$. 
After the  tachyon condensation, it should be projected into
a nontrivial projective module of $\cA_{\theta}$
which we investigate below. 

We would like to solve the equation of motion of the tachyon field by 
imposing several assumptions. Sufficient conditions are
\ba
\f{\de V(T)}{\de T}=0,\ \ \ [\nabla,T]=0.
\ea
The first equation is equivalent to the equation of motion if we take the
large $B$ limit as discussed in \cite{GoMiSt,HaKrLaMa,DaMuRa}. 
The solutions  to this equation are given by the projections
in the noncommutative torus algebra   $\cA_{\theta}$. 
The projections in $\cA_{\theta}$ are classified by 
the values of trace as in (\ref{prj})
and we write $T=P_{n-m\theta}$. 
Next we examine the second condition $[\nabla,T]=0$. This is satisfied\footnote
{One can also satisfy $[\nabla,T]=0$ if the connection acts only on the 
projective module
$E'=(1-P_{n-m\theta})E_{1,0}$. However a little analysis shows that
this is not consistent with the descent relation \cite{Se4} 
and does not have a
expected tension. Therefore we believe that this does not correspond to 
physical solutions and neglect these in this paper.}
if we use the connection
$\nabla_i=P_{n-m\theta}(\delta_i +A_i)P_{n-m\theta}$.
It defines the projective module  $E=P_{n-m\theta}E_{1,0}$
and the endomorphism 
 $\mbox{End}_{\cA_{\theta}}E$ is given by $P_{n-m\theta}\cA_{\theta}
 P_{n-m\theta}$. 
The projective module $E$ is twisted and its first Chern class (or cyclic
2-cocycle) is given by $\tau_2(P_{n-m\theta})=m$ 
as we explain in the appendix A. 
  
Now let us turn to the equation of motion for the 
gauge field $[\nabla_i,\hF_{ij}]=0$.
It is satisfied if the field strength is proportional to 
$P_{n-m\theta}$.
In a sense,   $P_{n-m\theta}$ can be
regarded as the identity in the algebra $\mbox{End}_{\cA_{\theta}}E$.
The field strength which is proportional to
the projector should be regarded as
constant curvature reviewed in the section 3.2. 

One may prove that for any projection of type
$P_{n-m\theta}$ the projective module of the form
$E=P_{n-m\theta}E_{1,0}=P_{n-m\theta}\cA_{\theta}$ has a constant 
curvature connection. As we saw in 
the subsection 3.1, for every $n,m$ one can construct 
a constant curvature connection
as the Heisenberg projective module. Because it is projective,
it should be written as $\ti{E}=\tilde{P}_{n-m\theta}\cA_{\theta}^N$ 
for a certain projection $\tilde{P}_{n-m\theta}$ in
$Mat_{N}(A_{\theta})$. Because any projection in $Mat_{N}(\cA_{\theta})$
for a given value of trace belongs to the same $K$-theory class \cite{Ri1}, we
can change $\tilde{P}_{n-m\theta}$ into any 
$P_{n-m\theta}\in \cA_{\theta}$ via an unitary
transformation. The transformed 
projective module $E=P_{n-m\theta}\cA_{\theta}$ also possesses the induced 
constant curvature connection. 
   
In this way we have found exact solutions of the equation of motion 
derived from (\ref{sn}). 
\ba
T=P_{n-m\theta},\ \ \ \hat{F}=
\f{1}{2\pi\al}\f{m}{n-m\theta}P_{n-m\theta}\ \J.
\label{soln}
\ea
Here we represents the fields as elements in the
`large' algebra $\cA_{\theta}$.
In the small algebra 
$\mbox{End}_{\cA_{\theta}}E$, both $T$ and
 $A_i$ are
proportional to the identity. 
Unlike the Moyal plane case, the small algebra is Morita equivalent
to $\cA_{\theta}$ and can be rewritten as
\ba
\mbox{End}_{\cA_{\theta}}E=P_{n-m\theta}\cA_{\theta}P_{n-m\theta}=
Mat_{d}(\cA_{\ti{\theta}}),
\ \ \ (\ti{\theta}=\f{b-a\theta}{n_0-m_0\theta}\ \ \mbox{s.t.}\ 
\ an_0-bm_0=-1),
\label{eme}
\ea
where we have defined $d=\mbox{g.c.d.}(n,m)$ and $(n,m)=d(n_0,m_0).$

We proceed to discuss what will be generated via the tachyon condensation 
(\ref{soln}). The mass of this excitation can be evaluated
by neglecting the derivative terms because of $[\nabla,T]=[\nabla,F]=0$,
\ba
M_{(n,m)}&=&\f{\s{2}}{\s{\al} G_s}\mbox{Tr}\left[P_{n-m\theta}
\s{\det(G+2\pi\al(\hat{F}+\Phi))}\right]\no
&=&\f{\s{2}|n|}{ \s{\al} g_s}\s{\det\left(g+2\pi\al(B+F)\right)},
\ea
where the flux $F$ is given by (\ref{flux}). The factor $\s{2}$ is
peculiar to non-BPS D-branes.
The above calculation can be 
done in the same way as in (\ref{BPSnm}). 
The dimension of the
projective module naturally appears here as the trace of the projection. 
It is interesting that the tachyon condensation 
on unstable D-brane systems gives an explicit physical realization of the 
mathematically fundamental relation between projective modules 
and projections.

If we assume $g_{ij}=R^2\delta_{ij}$ to make discussion clearer, 
the mass reduces to
\ba
M_{(n,m)}&=&\f{\s{2}}{\s{\al} g_s}\s{n^2R^2+(n/\theta-m)^2}.\label{massn}
\ea
It explicitly shows that the resulting state is a bound state
of $n$ non-BPS D2-branes and $(-m)$ non-BPS D0-branes. 
We comment that our arguments
of tachyon condensation naturally derive the fact that $-m$ or $n$ can be
negative which is consistent with the result in the previous subsection%
\footnote{One may ask the physical interpretation for the negative $n$.
Assume that $n$ is negative and thus $-m$ 
is positive. Since D0-branes with $B$-field on a torus generate 
(non-BPS) D2-brane, we can say that the negative $n$ 
means the annihilation of these induced non-BPS D2-branes with the $|n|$ 
non-BPS D2-branes. }. In this way we obtain all kinds of D2-D0 bound states
via tachyon condensation and thus our results are consistent with T-duality.

If we take the large $B/g$ limit,
the mass spectrum is proportional to $\dim(E)=n-m\theta$ 
as can be seen from (\ref{massn}) and it
 is dense in $\bf R$. 
This means that there exists a excitation of which energy is
arbitrary small. In \cite{BaKaMaTa} we
investigated the tachyon condensation in this limit and 
suggested that it leads to the instability. 
It means that any projection $P_{n-m\theta}$ can be divided into 
infinite numbers of mutually orthogonal smaller projections
while the total value of the  trace is preserved. This
was proved by investigating an explicit representation of projections. 

The argument can be simplified as follows.
We start from a projection $P_{n-m\theta}$  in $\cA_{\theta}$. 
It can be regarded as  the identity in the small algebra
$Mat_{d}(\cA_{\ti{\theta}})=P_{n-m\theta}\cA_{\theta}P_{n-m\theta}$. 
One can find another projection $Q$ in the small algebra.
The original projection is decomposed into two
mutually orthogonal projections,
\ba
P_{n-m\theta}=(1-Q)P_{n-m\theta}+QP_{n-m\theta}.
\ea
One may continue this operation repeatedly to give the
infinitely small mutually orthogonal projector.

We would like to claim that such instability does not
appear for finite $B/g$.
As can easily be seen from the mass formula (\ref{massn}), 
the bound state can be divided into $d=\mbox{g.c.d}\ (n,m)$ pieces 
but is not divided further. It means the instability 
cannot appear. We may interpret it from our exact tachyon solution.
The key point is the requirement of the constant field 
strength in (\ref{soln}) which is absent in large $B/g$ limit.
It is permitted to be divided into only $d$ mutually orthogonal parts, 
even though the tachyon field in (\ref{soln}) can be divided into 
infinitely many pieces. 
We conclude that for finite $B/g$ the bound states are 
all stable if  $\mbox{g.c.d}\ (n,m)=1$. 
 
The corresponding 
projective module $E_{n,m}=P_{n-m\theta}E_{1,0}$ 
can be written as a direct sum
of $d$ projective modules of the same type as we saw in section 3.1.
As shown in \cite{CoRi} (see also \cite{KoSc}), the moduli space of constant
curvature connection is given by the symmetric product of $d$ copies
of a two-torus  $({\bf T}^2)^d/S_d$.
This is actually the same as the physical moduli space 
of the solutions of (\ref{sn}) up to gauge transformation.
The freedom of unitary transformation of the tachyon field is absorbed 
in the gauge transformation $T\to UTU^*,\ U\in \cA_{\theta}$ and what
remains is only the  moduli space of constant 
curvature connection for the projective 
module $E_{n,m}=P_{n-m\theta}E_{1,0}$. 
We note that this is consistent with the physical intuition.
The moduli space of  a bound state for coprime 
$(n,m)$ parameterizes the transverse
coordinate for a D0-brane, namely ${\bf T}^2$. 
It is then obvious that the moduli space of the  bound states 
of $d$ D0-branes should be $({\bf T}^2)^d/S_d$.

The fluctuations of the gauge  and tachyon fields
around the solution (\ref{soln})  belong to
$Mat_{d}(\cA_{\ti{\theta}})$. This is physically interpreted as
the gauge group on the brane is $U(d)$.


Up to now we concentrate on the tachyon condensation
from a single D2-brane.
We may, of course, start from plural or even  infinitely many D2-branes.
In such situation, 
we can obtain arbitrary projective modules $E_{n,m}$ from
the projection in $Mat_{\infty}(\cA_{\theta})$.

It is also important to note the relation between tachyon 
condensation on 
the above noncommutative torus and that on the 
non-compact flat plane (Moyal plane). 
The exact solution for the latter \cite{HaKrLa} 
can be rewritten in our convention as flows,
\ba
T=P_{l},\ \ \ \hF=\f{1}{\Theta}P_{l}\ \J,\ \ \ 
(l\in {\bf Z}\geq 0)\label{solm}
\ea
where the noncommutativity $\Theta$ is defined as $[x^1,x^2]=i\Theta$.
The level-$l$ projection $P_{l}$ in the Moyal plane algebra is given by 
$\sum_{k=0}^{l-1}|k\lb\la k|$ and it corresponds to the generation
of $l$ D0-branes. To obtain such situation from the torus,
we need to take the large radius limit or equivalently the small
$\theta$ limit. In such a situation,
the value of $n$ is restricted to 0 or 1 
which is consistent with the result (\ref{solm}). 
We note that in this limit one cannot take $B/g\to\infty$ limit 
and thus the instability does not occur.


\vspace*{0.4cm}

\noindent{\bf Solution generating technique and Morita equivalence}

\vspace*{0.2cm}

We have seen the exact description of tachyon condensation
is characterized by the constant curvature connection.
It is interesting 
to ask what is the solution generating method which 
relates various solutions. 
In the Moyal plane, the exact solutions for tachyon condensation 
were constructed in \cite{HaKrLa} by using the shift operator 
$S=\sum_{n=0}^{\infty}|n+1\lb\la n|$. We would like to find
the analogous transformation on the two-torus $\cA_{\theta}$.



In this case  we have to be careful since the corresponding
operator in general interplotes the different $C^*$-algebras.
Namely after the tachyon condensation 
the original free module $E_{1,0}=\cA_{\theta}$ (a D2-brane) 
is changed into the twisted
projective module $E_{n,m}$ (a D2-D0 bound state).
As we have seen in section 2, the transformation
between these two solutions should be identified
with the Morita equivalence  bimodule $S$.
It depends on the integers $n,m$ and satisfies
\ba
E_{n,m}=S\otimes_{\cA_{\theta}}E_{1,0}.\label{gm}
\ea
This maps the endomorphism $\mbox{End}_{\cA_{\theta}} E_{1,0}=\cA_{\theta}$
into $\mbox{End}_{\cA_{\theta}} E_{n,m}=Mat_{d}(\cA_{\ti{\theta}})$. 
Physically this induces the transformation of the
world-volume field theories and this gives an explicit realization of the
descent relation \cite{Se4}.

In order to describe the exact solution, 
the projective module $E_{n,m}$ should have the constant curvature.
In other words, we have to impose
on the Morita equivalence $\cA_{\ti{\theta}}$-$\cA_{\theta}$ 
bimodule that $S$ should keep this additional constraint.
Actually it just fits the definition of
the {\it gauge Morita equivalence bimodule} discussed 
in \cite{KoSc,S}. 
We claim that this is the analogue of the shift operator
on the noncommutative torus.

We mention that the T-duality transformation can 
also be represented by the gauge 
Morita equivalence bimodules as argued in
\cite{S,PS,KoSc}. On a noncommutative torus $\cA_\theta$ 
there is a one-to-one correspondence between 
$\cA_\theta$-modules and the solutions found in (\ref{soln}). 
They correspond to the projective module 
$E_{n,m}$ and take their values in 
$\mbox{End}_{\cA_\theta}{E_{n,m}}$. 
If we perform T-duality so that the ($n$,$-m$) brane is transformed into 
 $d$ D2-branes (or equally applying the Morita equivalence 
 $\cA_{\theta}\sim \cA_{\ti{\theta}}$), 
 the projective module $E_{n,m}$ is changed into the free module 
 $\ti{E}_{d,0}$ in the algebra $\cA_{\ti{\theta}}$.
Such T-duality transformation can be constructed by 
$\cA_\theta$-$\cA_{\ti{\theta}}$ gauge Morita equivalence bimodule $X$
\ba
\ti{E}_{d,0}=E_{n,m}\otimes_{\cA_\theta} X\ . \label{gMequiv}
\ea   
The endomorphism $\mbox{End}_{\cA_\theta}{E_{n,m}}
=\mbox{End}_{\cA_{\ti\theta}}{\ti{E}_{d,0}}$ acts on $\ti{E}_{d,0}$ from the 
left in (\ref{gMequiv}). Especially, 
the solution $T={\bf 1}\in \mbox{End}_{\cA_\theta}{E_{n,m}}$ is the identity
in $\mbox{End}_{\cA_{\ti\theta}}{\ti{E}_{d,0}}$. 
On the other hand, 
$\hF=\ov{2\pi\alpha'}\f{m}{n-m\theta}\J
\in \Ed{\cA_\theta}{E_{n,m}}$ is translated to 
$0\in \Ed{\cA_{\ti\theta}}{\ti{E}_{d,0}}$, because 
$\ti{E}_{d,0}$ is the free module over $\cA_{\ti{\theta}}$. 
The difference between the values of the constant curvatures
comes from the constant curvature of $X$.
This is equivalent to the shift of the field 
$\Phi$ by T-duality as in (\ref{openT}).
These imply that for general noncommutative algebras the exact solution 
for tachyon condensation can be generated in the same way if there exist
gauge Morita equivalence bimodules.

\subsection{Tachyon condensation on brane-antibrane systems}
%
Brane-antibrane systems are more complicated and intriguing than 
non-BPS D-branes 
from the viewpoint not only of string theory but also of
the noncommutative geometry. The crucial difference from
non-BPS D-branes is that the tachyon field becomes
complex 
and belongs to a Morita equivalence
$\cA$-$\cB$ bimodule (we write this as $X$), where $\cA$ and $\cB$ are the 
algebras of the brane and the  antibrane as we discussed in section 2.
If we define $E$ and $F$ as the projective modules which 
represent the brane and the antibrane, respectively and define $\cA^{(0)}$ 
as the noncommutative base space, the algebras 
$\cA,\cB$ is given by $\cA=\mbox{End}_{\cA^{(0)}}E,\ 
\cB=\mbox{End}_{\cA^{(0)}}F$.
Remember that a Morita equivalence bimodule $X$ is defined by the 
bimodule which possesses two types
of inner product
 $\ \ \la\ ,\ \lb_{\cA},\ \  \la\ ,\ \lb_{\cB}$
and satisfies the conditions
(\ref{inner1},\ref{inner2},\ref{bimodule2}).

As in the previous subsection we use the 
example of noncommutative two-tori
and consider only D2-D0 bound states.
We assume the original brane-antibrane system is made of a $(n_1,-m_1)$
brane and a $(n_2,-m_2)$ antibrane. For simplicity we consider
only the case where the pairs of integers 
$(n_1,-m_1)$ and $(n_2,-m_2)$ are coprime. 
A pair of integers  $(n,-m)$ denote the indices of D2-D0 bound 
state or equivalently of those  the 
corresponding projective module $E_{n,m}$. 
Note that if one specifies
 $(n,m)$ such that $n-m\theta\geq 0$, then 
 there are two types of D-branes, that is,
  branes and antibrane. In our examples of the noncommutative torus
 the fundamental algebra $\cA^{(0)}$ is given by $\cA^{(0)}=\cA_{\theta}$, 
 where $\theta$ is represented in terms of 
 closed string variables as in (\ref{o-c}).
The algebras $\cA$ and $\cB$ are given by 
\ba
\cA&=&\cA_{\theta_1}, \ \ \theta_1=\f{b_1-a_1\theta}{n_1-m_1\theta},\ \ \
(a_1n_1-b_1m_1=-1)\no
\cB&=&\cA_{\theta_2}, \ \ \theta_2=\f{b_2-a_2\theta}{n_2-m_2\theta},\ \ \
(a_2n_2-b_2m_2=-1).
\ea
The tachyon field $T$ belongs to a $\cA_{\theta_1}$--$\cA_{\theta_2}$ 
bimodule. There are also the gauge fields on the brane and the antibrane. We 
denote these as $A^{(1)}$ and $A^{(2)}$. These fields belong to 
$\cA=\cA_{\theta_1}$ and $\cB=\cA_{\theta_2}$, respectively. 
The covariant derivative of the tachyon field is given 
by the connection for a bimodule $X$
(see for example \cite{S,KoSc}) specified by the requirement
\ba
\nabla_X (a T)=\delta_{\cA}(a)T+a\nabla_X T\ \ \   (\forall a\in \cA),\no
\nabla_X (T b)=(\nabla_X T)b+T\delta_{\cB}(b)\ \ \   (\forall b\in \cB),
\ea
where $\delta_{\cA}$ and $\delta_{\cB}$ denote the derivation in $\cA$ and
$\cB$, respectively. 
We also use the covariant derivative of 
field strengths $\hF^{(1)}$ and $\hF^{(2)}$. Each of them is given 
by the commutator with a connection for the algebra $\cA$ or $\cB$ 
as in the previous subsection.

Now we have prepared to discuss the tachyon condensation on 
noncommutative tori. 
The effective action for a 
D2$-\overline{\mbox{D}2}$ with two 
gauge fluxes was already computed in \cite{TaTeUe}
using the boundary string field theory \cite{Wi2,GeSh,KuMaMo}.
Applying this to our system on a noncommutative torus the result 
is given by 
\ba
S&=&\f{1}{\s{\al} G_s}\int dt\mbox{Tr}_{\cA}\left[V\left(\la T,T\lb_{\cA}
\right)
\s{\det(G+2\pi\al(\hF^{(1)}+\Phi))}\right]\no 
&+&\f{1}{\s{\al} G_s}
\int dt\mbox{Tr}_{\cB}\left[V\left(\la T,T\lb_{\cB}\right)
\s{\det(G+2\pi\al(\hF^{(2)}+\Phi))}
\right]\no
&+&O(\nabla_X T,\ [\nabla_{\cA}, \hF^{(1)}],\ [\nabla_{\cB}, \hF^{(2)}]),
\label{sba}
\ea
where $O(\nabla_{X} T,\ddd)$ denotes the derivative terms. 
We defined the traces for $\cA$ and $\cB$ by embedding these 
algebras in  $Mat_{N}(\cA^{(0)})$ for a sufficient large integer $N$. 
These satisfy the following relation \cite{Ri1,Ri2}
\ba
\mbox{Tr}_{\cA}\la T_1,T_2\lb =\mbox{Tr}_{\cB}\la T_2,T_1\lb, \label{tr}
\ea
and are normalized by $\mbox{Tr}_{Mat_{N}(\cA^{(0)})}1=N$.

We would like to solve the equation of motion for the action (\ref{sba}).
Below we give solutions by imposing ansatz similar to the
previous subsection. 
We assume the existence of the partial isometry-like equation 
\cite{Wi3,HaKrLa,HiNoTa} for the tachyon field (\ref{piso})
\ba
\la T,T \lb_{\cA} T=T,\ \ \ T \la T,T \lb_{\cB}=T.\label{tak}
\ea
Note that these two equations are equivalent thanks to the relation 
(\ref{bimodule2}). The solutions to this equation give the stationary points
of the tachyon potential $V\left(\la T,T\lb_{\cA}
\right)$ and $V\left(\la T,T\lb_{\cB}\right)$. To make exact solutions for
finite $B/g$, we should take account of the gauge fields. It is easy to see
the equation of motions for the tachyon $T$ and gauge fields 
 $(A^{(1)},\ A^{(2)})$ are satisfied if we require (\ref{tak}) and 
\ba
&&\nabla_{X} T=0,\label{dt}   \\
&& [\nabla_{\cA}, \hF^{(1)}]=0,\ \ [\nabla_{\cB}, \hF^{(2)}]=0 \label{df},
\ea
are satisfied.
Since (\ref{tak}) is equivalent to the statement that 
$\la T,T \lb_{\cA}\in \cA$ and $\la T,T \lb_{\cB}\in \cB$ are both projections
as explained in section 2, we can write these as follows
\ba
\la T,T \lb_{\cA}=1-P_{\ap+\bt\theta_1}(\equiv 1-P_1), \ \ \
\la T,T \lb_{\cB}=1-P_{\ga+\dt\theta_2}(\equiv 1-P_2). \label{takpro}
\ea
Using the relation (\ref{tr}) we obtain the constraint,
\ba
n_1(1-\ap)-b_1\bt=n_2(1-\ga)-b_2\dt,\ 
\ \ m_1(1-\ap)-a_1\bt=m_2(1-\ga)-a_2\dt.
\ea
They determine $\ga$ and $\dt$ in terms of 
$\ap$ and $\bt$.
The tachyon field which condensates as in 
(\ref{takpro}) belongs to the bimodule $(1-P_1)\cdot X\cdot (1-P_2)$.
For this tachyon field the potential is evaluated as
\ba
V\left(\la T,T\lb_{\cA}
\right)=P_{\ap+\bt\theta_1},\ \ \ 
V\left(\la T,T\lb_{\cB}\right)=P_{\ga+\dt\theta_2}. \label{tpr}
\ea
This is because we use the convention that the
original brane-antibrane system corresponds to $T=0$ and that 
the vanishing of brane and antibrane
corresponds to $\la T,T\lb_{\cA}=1$ and $\la T,T\lb_{\cB}=1$, 
respectively\footnote{
Notice that this is the opposite convention to the tachyon potential of 
non-BPS D-branes in section 3.3.}.

Let us turn to the next equation (\ref{dt}). This is satisfied if the gauge
fields belong to
\ba
A^{(1)}\in P_1\cA P_1,\ \ \ A^{(2)}\in P_2\cB P_2.
\ea
The last equation (\ref{df}) is solved if we assume that both of the 
gauge fields have constant curvatures
\ba
\hat{F}^{(1)}&=&\f{1}{2\pi\al}\f{\beta a_1+m_1\ap}
{n_1\ap+b_1\beta-(\beta a_1+m_1\ap)\theta}{\bf 1}\in P_1\cA P_1,
\no
\hat{F}^{(2)}&=&\f{1}{2\pi\al}\f{\delta a_2+m_2\ga}
{n_2\ga+b_2\dt-(\dt a_2+m_2\ga)\theta}{\bf 1}\in P_2\cB P_2.
\ea
It finishes our calculation of the physical solutions of the
tachyon condensation.
Our result does not depend on the detailed
form of the derivative term because of (\ref{dt},\ref{df}).

We evaluate mass spectrum of these classical solutions, 
\ba
M&=&\f{1}{\s{\al}G_s}\mbox{Tr}_{\cA}\left[P_{\ap+\bt\theta_1}
\s{\det(G+2\pi\al(\hat{F}^{(1)}+\Phi))}\right]\no
&+&\f{1}{\s{\al}G_s}
\mbox{Tr}_{\cB}\left[P_{\ga+\dt\theta_2}
\s{\det(G+2\pi\al(\hat{F}^{(2)}+\Phi))}\right]\no
&=&\f{|N_1|}{\s{\al}g_s}\s{\det(g+2\pi\al (B+F^{(1)}))}
+\f{|N_2|}{\s{\al}g_s}\s{\det(g+2\pi\al(B+F^{(2)}))}.
\ea
We have defined the fluxes as follows
\ba
F^{(1)}=\f{\J}{2\pi\al}\f{M_1}{N_1},\ \ \ 
F^{(2)}=\f{\J}{2\pi\al}\f{M_2}{N_2},
\ea
where integers $M_1$, $N_1$, $M_2$ and $N_2$ are given by
\ba
&& N_1=\ap n_1+\bt b_1,\ \ \ M_1=\bt a_1+\ap m_1 \no
&& N_2=\ga n_2+\dt b_2,\ \ \ M_2=\dt a_2+\ga m_2.
\ea

We find that the products of the tachyon condensation are 
identified with a $(N_1,-M_1)$ brane and a $(N_2,-M_2)$ anti-brane. 
If these integers are not coprime, each 
bound state can be divided into several parts as before. 
Note that what are produced after the tachyon condensation
depend on the projections (\ref{tpr}). The original brane-antibrane system
corresponds to $\ap=\ga=1,\ \bt=\dt=0$. If one assumes that the dimension of 
the projective module $E$ is larger than that of $F$, then the tachyon 
field $\ga=0,\dt=0$  gives
the maximal condensation and this will produce a $(n_1-n_2,-m_1+m_2)$ 
brane. In the opposite case the 
tachyon field $\ap=0,\ \bt=0$
will generate a $(n_2-n_1,-m_2+m_1)$ anti-brane.

For general decay modes
the differences of the D2-brane charge $(N_1-N_2)$ and the 
D0-brane charge $(M_2-M_1)$ are preserved as follows
\ba
N_1-N_2=n_1-n_2,\ \ \ M_2-M_1=m_2-m_1. \label{rc}
\ea

The charge conservation can also be discussed in the framework of 
operator algebra $K_0$-group $K_0(\cA_{\theta})$. 
If one would like to consider the $K_0$-group 
of noncommutative torus, the Chern character
\cite{Co} gives enough information \cite{Ri2}. 
In the brane-antibrane system which corresponds to
the pair of projective modules $(E,F)\in K_0(\cA_{\theta})$ the $K$-theory
 charge is given by the difference
\ba
\mbox{ch}(E)-\mbox{ch}(F)&=&\mbox{Tr}_{\cA}
\exp(2\pi\al \hF^{(1)})-\mbox{Tr}_{\cB}
\exp(2\pi\al \hF^{(2)})\no
&=&(n_1-m_1\theta-n_2+m_2\theta)+(m_1-m_2)dx^1dx^2,
\ea
where $dx^1dx^2$ is the two form along the two-torus. 
It is known that the 
RR-couplings on a brane-antibrane system can be written by using 
$K^0$-type superconnection \cite{KeWi,KrLa,TaTeUe}. Applying this
idea to our examples we obtain the following RR-couplings
\ba
S_{RR}\sim \int C_{RR}\we \left[\mbox{Tr}_{\cA}V\left(\la T,T\lb_{\cA}
\right)\exp(2\pi\al \hF^{(1)})-\mbox{Tr}_{\cB}V\left(\la T,T\lb_{\cB}
\right)\exp(2\pi\al \hF^{(2)})\right].
\ea
Note that in our noncommutative description the derivative
of tachyon field is always zero and does not contribute. Also notice that
 the potential in the above can be regarded as the 
identities in the algebras of 
 gauge fields $\mbox{End}_{\cA{_\theta}} E$ and 
 $\mbox{End}_{\cA{_\theta}} F$ because
of (\ref{tpr}). Thus the conservation of RR-charge (\ref{rc}) is equal
to that of operator algebra $K$-theory charge and therefore 
this gives a further support to
the relation between D-brane charge and $K$-theory \cite{MiMo,Wi,Wi4}.
The intriguing characteristic that 
the above `noncommutative Chern character' depends on
$\theta$ will correspond to the physical fact that D0-branes in the $B$-field
background generate D2-brane charge \cite{MaIc}. Therefore it will also be 
interesting to clarify the relation between 
 the RR-couplings for
non-abelian transverse scalars \cite{My,TaTeUe} 
and the above RR-couplings on various noncommutative
 tori \cite{MaIc} (see also \cite{Mukhi} for the Moyal plane).

\section{Conclusions}
\hspace{5mm}
We first discussed the open string theory in general noncommutative
background.
We considered a general framework to handle open strings 
and D-branes in a unified way by utilizing the Morita equivalence. 
In particular we proposed the equation which defines noncommutative solitons
on general brane-antibrane systems.

{}From this viewpoint
we have examined the exact solutions on noncommutative tori in tachyonic 
systems. For non-BPS branes the solutions are given by the tachyon field
which is proportional to the projection and the gauge field with a constant
curvature. This respects the one-to-one correspondence between
a projection and a projective module. We have also shown
that these solutions
can be generated by employing the (gauge) Morita equivalence. 
Our exact description of tachyon condensation including the gauge field 
solves for finite $B/g$ the previously observed instability problem.

More complicated and thus more intriguing examples are brane-antibrane
systems. In this case the tachyon field belongs to the Morita equivalence 
bimodule and we can impose the partial isometry-like relation instead of
the equation of motion. We can construct the exact solutions and 
determine the decay products. We find the RR-charges of brane-antibrane
systems can be represented by the superconnection-like extension
of the Connes's Chern character and check that these charges 
conserved in the process of tachyon condensation. This also verifies the 
fact that the D-brane charge is classified by the operator algebra $K$-theory.

\bigskip \noindent \textbf{Note added:} After completing our calculations, 
we noticed the preprint \cite{KaSc} on the net which has some overlaps 
with our results in section 3.3.
\bigskip

\begin{center}
\noindent{\large \textbf{Acknowledgments}}
\end{center}

We are grateful to M. Hamanaka, K. Ichikawa, A. Kato, K. Ohmori, Y. Sugawara, 
S. Terashima and T. Uesugi for useful discussions. We are also happy to
thank H. Moriyoshi for his clear lecture on noncommutative geometry and
 useful comments. H.K. and T.T. 
are supported by JSPS Research Fellowships for Young Scientists. 
Y.M. is supported in
part by Grant-in-Aid (\# 707) from the Ministry of Education, Science,
Sports and Culture of Japan. 


\appendix
\setcounter{equation}{0}
\section{Powers-Rieffel Projections and Cyclic Cocyles}
\hspace{5mm}
Here we give a review of Powers-Rieffel projections \cite{Ri1} on 
two dimensional 
noncommutative tori and of the calculations of their topological charges 
\cite{Co,co-book}. 
We also 
mention some other projections constructed in 
\cite{BaKaMaTa}.

First let us consider projections in any $C^*$-algebra $\cA$ and assume that 
 there exist a trace ${\mbox{Tr}}: \cA\to {\bf C}$ and derivations 
 $\delta_{i}:\cA\to \cA$ such that 
 $\delta_{i}(ab)=\delta_{i}(a)b+a\delta_{i}(b)$.
  We also normalize the trace as $\mbox{Tr}({\bf 1})=1$.
  The 
 index $i$ corresponds to the basis of the derivations\footnote{
 More precisely, the derivation is defined as an action of a Lie group $G$ 
 on
 the algebra $\cA$ \cite{Co}. 
 Then the basis of the derivation can be said as those of 
 the
 Lie algebra.}.
A projection $p\in \cA$ is defined to be a self-adjoint idempotent $p^*=p=p^2$.
The Connes's Chern Character is defined as the 
exponential of the curvature $F\in \mbox{End}_{\cA} E$ of projective module 
$E$ \cite{Co}:
\ba
\mbox{ch}(E)=\mbox{Tr}
 \exp{\big(\f{F}{2\pi i}\big)}=\sum_{k=0}^{\infty}\f{1}{k!}
\ \tau_{2k}(F,F,\ddd,F). \label{ch}
\ea
The each term $\tau_{2k}(F,F,\ddd,F)$ of the above expansion represents 
 the contribution which is proportional to $F^{k}$.
 Note that $\tau_0$ is equal to the trace of 
identity and it gives the dimension $\mbox{dim}(E)$ of the projective module.

Since 
any projective module $E$ is described as $P\cdot \cA^N$ using a 
projection $P$ in $Mat_{N}(\cA)$ for a sufficiently large integer $N$, 
one can rewrite 
the curvature in terms of the projection. More explicitly one can 
choose a connection 
of $E$ as $\nabla_{i}=P\cdot\delta_{i}\cdot P$ \cite{Co,KoSc}. 
Note that the topological 
quantity such as the 
Chern character does not depend on the choice of the connection.
Then the curvature $F\in \mbox{End}_\cA E$ is expressed as
\ba
F_{ij}=[P\delta_{i}P,\ P\delta_{j}P]=
P(\delta_{i}P)(\delta_{j}P)-P(\delta_{i}P)(\delta_{j}P)+P(\delta_{i}\delta_{j}
-\delta_{j}\delta_{i})P.
\ea
Because the trace in $\mbox{End}_{\cA} E=P\cdot\mbox{Mat}_{N}(\cA)\cdot P$ 
is naturally induced from
 the trace in $\mbox{Mat}_{N}(\cA)$ normalized as $\mbox{Tr}({\bf 1})=N$, 
 one can always
 calculate the Chern character (\ref{ch}) if the projection $P$ is given. Thus
 we write the $2k$-th part by $\tau_{2k}(P)$. This can be regarded as the
  cyclic $2k$-cocycle where the projections are substituted.
 
 For example, the cyclic 0-cocycle is given by
\ba
\tau_0(P)=\mbox{Tr}(P)=\mbox{dim}(E).
\ea

Now let us investigate the explicit examples of cyclic cocycles. 
First we consider
 a flat two dimensional plane (Moyal plane) algebra. We employ the operator 
 representation
 and define the noncommutative coordinate $(x^1,x^2)$ as $[x^1,x^2]=i\Theta$. 
 Further
 we define the creation and annihilation operator 
 $a^{\dagger}=\f{1}{\s{2\Theta}}(x^1-ix^2)$ and 
 $a=\f{1}{\s{2\Theta}}(x^1+ix^2)$ such that they satisfy $[a,a^{\dagger}]=1$.
Using this one can define the basis of the Hilbert space as the 
familiar $n$-number state $|n\lb=\f{1}{\s{n!}}(a^{\dagger})^n|0\lb$. 
 Then the algebra is expanded by $|n\lb\la m| \ \ (n,m\geq 0)$. 
 Let us consider the projection $P_{n}=\sum_{k=0}^{n-1}|k\lb\la k|$ for
  a finite integer $n$.
 The trace of this operator is given by $\mbox{Tr}(P_{n})=2\pi\Theta\cdot n$,
 where we have normalized the trace so that it is equal to the integration 
 $\int dx^1dx^2$ in the c-function representation. Now it is straightforward 
 to
 calculate $\tau_2(P_{n})$ since the derivations $\delta_1,\delta_2$ along 
 the coordinate $x^1,x^2$
 are given by
\ba
\delta_1=\f{i}{\Theta}[x^2,\ ],\ \ \ \delta_2=-\f{i}{\Theta}[x^1,\ ].
\ea
Thus we obtain the result as follows
\ba
\tau_2(P_{n})&=&\f{1}{2\pi\Theta}\mbox{Tr}\left
[P_{n}[a^{\dagger},P_{n}][a,P_{n}]
-P_{n}[a,P_{n}][a^{\dagger},P_{n}]\right]\no
&=&\f{1}{2\pi\Theta}\mbox{Tr}\left
[P_{n}+P_{n}a^{\dagger}P_{n}a P_{n}
-P_{n}a P_{n}a^{\dagger}P_{n}\right]\no
&=&n.
\ea
Therefore the value of $\tau_2(P_{n})$ is quantized and is positive. 
Note that this value is equal to the first Chern class of the projective 
module and therefore should be quantized. Then one may
 ask if one can obtain negative integers ? The answer is yes and 
 one can construct the corresponding projection as $1-P_{n}$. In this case we 
 obtain the value $\tau_2(1-P_{n})=-n$.

After this elementary example, let us turn to the two dimensional 
noncommutative 
torus $\cA_{\theta}$. We assume $\theta$ is irrational because
 for rational $\theta$ there are finite dimensional representations
  of the algebra and the calculations are simplified 
 (see also \cite{Bo,BaKaMaTa,MaMo,GoHeSp}).

As we have mentioned in section 3.1, the projections in $\cA_{\theta}$ are 
generally 
characterized by their values of trace as in eq.(\ref{prj}).
The explicit construction of projections (Powers-Rieffel projection) 
was given in \cite{Ri1} and let us 
review this below.

Because $U_1$ and $U_2$ do not commute ($U_1U_2=U_2U_1e^{2\pi i\theta}$), 
we can 
 diagonalize only $U_2$ and define c-number $x^2$ as $U_2=e^{2\pi i x^2}$.
Thus we obtain the following (infinite dimensional) representation of 
$\cA_{\theta}$ 
\ba
U_1|x^2\lb=|x^2+\theta \lb,\\   U_2|x^2\lb=e^{2\pi ix^2}|x^2\lb.
\ea
Then the trace of an element $a\in \cA_{\theta}$ is given by
\ba
\mbox{Tr}(a)=\int_{0}^{1}dx^2 \la x^2|a| x^2 \lb.
\ea

In order to find explicit projections $P$ we assume the following form
\begin{equation}
P=U_1^{*}\left( {g}(U_2)\right) ^{*}+f(U_2)+g(U_2)U_1.
\label{PR1}
\end{equation}
As we will see below, one can construct a projection 
for each $n,m$ even under this restriction.
Acting on the position space $|x^2\lb$, we require 
$P^{2}|x^2\lb=P|x^2\lb$.  This defines a projection in 
$\cA_{\theta }$ if and only if $f$ and $g$ satisfy the following
relations
\begin{eqnarray}
&&g(e^{2\pi ix^2})g(e^{2\pi i(x^2+\theta )})=0\,\,,  \nonumber \\
&&g(e^{2\pi ix^2})[1-f(e^{2\pi ix^2})-f(e^{2\pi i(x^2+\theta
)})]=0\,\,,  \nonumber \\
&&f(e^{2\pi ix^2})[1-f(e^{2\pi ix^2})]=|g(e^{2\pi
ix^2})|^{2}+|g(e^{2\pi i(x^2-\theta )})|^{2}\ .  \label{PR}
\end{eqnarray}

Explicit forms of $f,g$ which satisfy these relations are given as
follows. Choose any small $\epsilon >0$ such that $\epsilon <\theta $ and $%
\theta +\epsilon <1$, and let $F(x^2)\equiv f(e^{2\pi ix^2})$ for one
period be given in the range $x^2\in \left[ 0,1\right] $ by
\begin{equation}
F(x^2)=\left\{
\begin{array}{lcl}
x^2/\epsilon  & \qquad  & x^2\in \lbrack 0,\epsilon ] \\
1 &  & x^2\in \lbrack \epsilon ,\theta ] \\
1-(x^2-\theta )/\epsilon  &  & x^2\in \lbrack \theta ,\theta +\epsilon ]
\\
0 &  & x^2\in \lbrack \theta +\epsilon ,1]
\end{array}
\right. \,\,,  \label{F}
\end{equation}
Then define $g$ for one period by
\begin{equation}
g(e^{2\pi ix^2})=
\begin{cases}
\sqrt{F(x^2)(1-F(x^2))} & \ \ x^2\in \lbrack
0,\epsilon ], \\
0 & \ \ x^2\in \lbrack \epsilon ,1]\ .
\end{cases}
\end{equation}
It is easy to see that the functions $f$ and $g$, 
defined as the periodic extensions of the above,
satisfy the relation (\ref{PR}). It can be easily shown that
\begin{equation}
\mbox{Tr}\,P=
\int_{0}^{1}dx^2<x^2|P|x^2>
=\int_{0}^{1}dx^2F(x^2)=\theta \,\,.
\end{equation}
Thus the projection $P$ now constructed corresponds to $P_{\theta}$. 

Now how about more general projections $P_{n-m\theta}$ ? Such general 
projections can be constructed by slightly modifying the
above constructed $P_{\theta}$ as follows. 
The general projection $P_{n-m\theta}$ can be regarded as the projection
$P_{\theta '}$ in the algebra $\cA_{\theta '}$ 
if we define $\theta '=n-m\theta$. It is easy to see that the algebra 
$\cA_{\theta '}$ can be 
embedded in $\cA_{\theta}$ by replacing $(U_1,U_2)$ with
 (i) $(U_1,U_2^m)$ or (ii) $
(U_1^m,U_2)$. Since one can construct the projection $P_{\theta '}$ in 
the previous way, we obtain the projection $P_{n-m\theta}$ in 
$\cA_{\theta}$ as desired. 

In the first choice (i), the projection
is described by functions $f$ and $g$ with period $1/|m|$ 
and the width of each lump of 
the function $f$ 
is given by $(n-m\theta)/|m|$. 
This preserves the form (\ref{PR1})
 and is called Powers-Rieffel projection \cite{Ri1}. On the other hand, 
 in the second
choice (ii) the requirement for being a 
projection is given by the equation 
(\ref{PR})
 with $\theta$ replaced by $m\theta$ and this is not included in 
 the form (\ref{PR1}). Then the width of the lump 
 is enlarged to 
$n-m\theta$. This construction was given in \cite{BaKaMaTa} and 
used in the proof that any projection can be
 divided into infinite numbers of mutually orthogonal smaller 
projections, where the total value
  of trace is preserved. Note also that in either case, 
  the total area occupied by the lump
is $0\leq n-m\theta\leq 1$.


Next we turn to the calculation of cyclic 2-cocyle $\tau_2(P)$. 
Define the derivations $\delta_{1},\delta_{2}$ along the two directions
 of the 
two
 dimensional torus as follows
\ba
\delta_j U_k=2\pi i \delta_{jk}U_k,
\ea
where $\delta_{ij}$ is the ordinary Kronecker's delta. Equivalently,
 one can express the derivations as follows by using the 
 noncommutative coordinate $(x^1,x^2)$ defined by $U_1=e^{2\pi i x^1},\ 
 U_2=e^{2\pi i x^2}$:
\ba
\delta_1=-i\f{2\pi}{\theta}[x^2,\ ],\ \ \ \delta_2=i\f{2\pi}{\theta}[x^1,\ ].
\ea

{}From these we can see that $\delta_1$ and $\delta_2$ do commute.
Then the cyclic 2-cocyle is defined by
\ba
\tau_2(a,b,c)=\f{1}{2\pi i}\mbox{Tr}\left[a\ \delta_{1}(b)\delta_{2}(c)-
a\ \delta_{2}(b)\delta_{1}(c)\right].
\ea
If we substitute $a=b=c=P_{\theta}$, then we get \cite{Co}
\ba
&&\tau_2(P_{\theta})\no
&&=-(4\pi i)\mbox{Tr}\bigl[f'(U_2)(g(U_2))^2-f(U_2)g'(U_2)g(U_2)U_2
 \no
&&\ \ \ \ \ 
+U_1f(U_2)U_1^*g'(U_2)g(U_2)U_2-(g(U_2))^2U_1f'(U_2)U_2U_1^*\bigr]\no
&&=-\int^{1}_{0}dx^2(f(x^2+\theta)-f(x^2))\f{d}{dx^2}(g(x^2)^2)
+2\int^{1}_{0}dx^2\f{d}{dx^2}(f(x^2+\theta)-f(x^2))(g(x^2)^2)\no
&&=-6\int^{1}_{0}dx^2 \f{df(x^2)}{dx^2}(g(x^2)^2)\no
&&=-6\int^{1}_{0}\! dt\ (t-t^2)=-1.
\ea
It is also possible to generalize the result for the projections 
$P_{n-m\theta}$ and
 we obtain
\ba
\tau_2(P_{n-m\theta})=m.
\ea
In order to see this one has only to note that for the description (i) 
the evaluation of $\tau_2$
 is equal to counting of the number of lumps with sign 
 and also that for the description (ii) the factor $m$ is due to the 
 derivation of $U_1^m$. Indeed this value of $\tau_2$ is
 the same as that computed from the previously discussed projective modules 
 $E_{n,m}\ \ (0\leq n-m\theta\leq 1)$ which possess the constant curvature 
 $F=\f{2\pi im}{n-m\theta}{\bf 1}\in \mbox{End}_{\cA_{\theta}} 
 E_{n,m}$ as follows
\ba
\tau_2(F)=\f{1}{2\pi i}\mbox{Tr}(F)=m.
\ea

Finally let us discuss the relation between the results in the Moyal plane 
and those in noncommutative torus. Since the radius of the torus is scaled
in proportion to $\f{1}{\s{\theta}}$, the noncommutative torus will 
approaches the Moyal plane in the limit $\theta\to 0$. In this limit the 
value of integers $n,m$ which satisfy $0\leq n-m\theta\leq 1$ is 
restricted to $n=0$ or $n=1$. Thus we obtain the projection $P_{m\theta}$ and
$P_{1-m\theta}$. This is consistent with the previous result that
 in the Moyal plane algebra the projection is given by $P_{m}$ or $1-P_{m}$ 
 up to unitary equivalence.

\end{document}